\documentclass[conference]{IEEEtran}
\usepackage{graphicx}
\usepackage{amssymb}
\usepackage{amsthm}
\usepackage{amsmath}
\usepackage{caption2}

\ifCLASSINFOpdf
\else
\fi

\begin{document}
%
\title{Imperfect and Unmatched CSIT is Still Useful for the Frequency Correlated MISO Broadcast Channel}


\author{\IEEEauthorblockN{Chenxi Hao and Bruno Clerckx}
\IEEEauthorblockA{Communication and Signal Processing Group, Department of Electrical and Electronic Engineering\\
Imperial College London, United Kingdom\\
Email: \{chenxi.hao10,b.clerckx\}@imperial.ac.uk}}

\maketitle

\begin{abstract}
Since Maddah-Ali and Tse showed that the completely stale transmitter-side channel state information (CSIT) still
benefits the Degrees of Freedom (DoF) of the Multiple-Input-Multiple-Output (MISO) Broadcast Channel (BC), there has been much interest in the academic literature to investigate the impact of imperfect CSIT on \emph{DoF} region of time correlated broadcast channel. Even though the research focus has been on time correlated channels so far, a similar but different problem concerns the frequency correlated channels. Indeed, the imperfect CSIT also impacts the DoF region of frequency correlated channels, as exemplified by current multi-carrier  wireless systems.

This contribution, for the first time in the literature,
investigates a general frequency correlated setting where a two-antenna transmitter has imperfect
knowledge of CSI of two single-antenna users on two adjacent subbands.
A new scheme is derived as an integration of Zero-Forcing Beamforming (ZFBF) and the scheme proposed by Maddah-Ali and Tse. The  achievable DoF region resulted by this scheme is expressed as a function of the qualities of CSIT. \footnote{The project HARP acknowledges the financial support of the Seventh Framework Program for Research of the European Commission under grant number: 318489}
\end{abstract}


\IEEEpeerreviewmaketitle

\section{Introduction}

In downlink multi-user multiple-input-multiple-output (MU-MIMO) communications, the latency and inaccurate CSIT degrade the DoF when conventional precoding techniques such as ZFBF are employed. Strategies to exploit imperfect feedback to enhance DoF region has therefore attracted a lot of attention. The completely stale CSIT was first studied by Maddah-Ali and Tse. In their contribution \cite{Tse10}, an optimal per-user \emph{DoF} of $\frac{2}{3}$ in a two-user setup was
achieved by a simple transmission scheme (denoted as MAT scheme in
the sequel). That result was extended later in \cite{Ges12} and
\cite{Gou12}, by accounting for imperfect current CSIT. The optimal DoF was derived and expressed as a function of the quality of the current CSIT. The achievablility was shown using a scheme that bridges the \emph{DoF} found in \cite{Tse10} and \cite{Ges12}\cite{Gou12}.

In \cite{Chen12a} and \cite{CB12}, unequal quality of current CSIT per user was investigated. The optimal bound found is a superset of the results in \cite{Ges12} and \cite{Gou12}, revealing that an asymmetric \emph{DoF} is achieved by each user. Moreover, \cite{Chen12b} has studied the imperfect delayed CSIT, suggesting that it can be used as good as perfect delayed CSIT.

To date, all the works only focus on the time correlated
channel. However, the DoF region of frequency correlated channels is also impacted by the imperfect CSIT. In current multi-carrier communication systems,  the CSIT is measured and reported by users on a per-subband basis. In practice, each user only reports its CSI on a group of predefined subbands, which might provide few information about the channel of the subbands outside the group because of the weak correlation between different subbands. In this paper, we investigate, for the first time in the literature, the \emph{DoF} of a general two-subband based frequency correlated broadcast channel with arbitrary imperfect CSIT (see Section \ref{system_model}). Our contributions are summarized as follows:
\begin{enumerate}
\item
Derive an achievable DoF of a two-user and two-subband based scenario as a function of the quality of the CSIT,
\item
Design a novel transmission strategy, motivated by MAT and ZFBF, that achieves the DoF region.
\end{enumerate}

The rest of this paper is organized as follows. The system model is
introduced in Section \ref{system_model} and the achievable \emph{DoF} region is given in Section \ref{DoF}. The \emph{DoF} achieved via reusing MAT and ZFBF is identified in Section
\ref{achievability} and a novel transmission scheme is introduced. Section \ref{conclusions}
concludes the paper.

The following notations are used throughout the paper. Bold lower letters stand for
vectors whereas a symbol not in bold font represents a scalar.
$\left({\cdot}\right)^T$ and $\left({\cdot}\right)^H$ represent the
transpose and conjugate transpose of a matrix or vector
respectively. $\mathbf{h}^\bot$ denotes the orthogonal space of the
channel vector $\mathbf{h}$. $\mathcal{E}\left[{\cdot}\right]$
refers to the expectation of a random variable, vector or matrix.
$\parallel{\cdot}\parallel$ is the norm of a vector.
$\left|{\cdot}\right|$ represents the magnitude of a scalar.
$f\left(P\right){\sim}{P^{B}}$ corresponds to
${\lim}_{P{\to}{\infty}}\frac{{\log}f\left(P\right)}{{\log}P}{=}B$,
where $P$ is supposed to be the SNR throughout the paper and
logarithms are in base $2$. $P_a$ denotes the power of $a$ while
$R_a^{\left(1\right)}$ and $R_a^{\left(2\right)}$ represent the
rate of $a$ achieved at receiver 1 and 2 respectively.

\section{System Model}\label{system_model}

We consider a two-user broadcast channel with two transmit antennas
and one antenna per user. The related parameters are defined as follows. $\mathbf{h}_i$ and $\mathbf{g}_i$ are the
channel states in subband $i$ of user 1 and user 2 respectively. Denoting
the transmit signal vector  in subband $i$ as $\mathbf{s}_i$, subject to a per-subband based power constraint $\mathcal{E}\big[\left\|\mathbf{s}_i\right\|^2\big]{\sim} P$, the
observations at receiver 1 and 2, $y_i$ and $z_i$ respectively, can
be written as
\begin{align}
y_i&=\mathbf{h}_i^H \mathbf{s}_i + \epsilon_{i,y},\label{eq:model1}\\
z_i&=\mathbf{g}_i^H \mathbf{s}_i + \epsilon_{i,z},\label{eq:model2}
\end{align}where $\epsilon_{i,y}$ and $\epsilon_{i,z}$ are unit power AWGN noise.
Signal vector $\mathbf{s}_i$ is a function of the symbol vectors for
user 1 and user 2, denoted as $\mathbf{u}_i$ and $\mathbf{v}_i$
respectively. $\mathbf{u}_i$ is a two-element symbol vector
containing $u_{i,1}$ and $u_{i,2}$. Similarly, $\mathbf{v}_i$ is
defined.

The channels are characterized as follows. $\mathbf{h}_i$ and $\mathbf{g}_i$ are mutually independent and identically distributed with
zero mean and unit covariance matrix
($\mathcal{E}\left[|\mathbf{h}_i^H\mathbf{g}_i|^2\right]{=}0$ and
$\mathcal{E}\left[\mathbf{h}_i\mathbf{h}_i^H\right]{=}\mathbf{I}_2$).
The imperfect CSIT of user 1 is denoted as $\hat{\mathbf{h}}_i$
while the imperfect CSIT of user 2 is $\hat{\mathbf{g}}_i$, each
with the error vector of
$\tilde{\mathbf{h}}_i{=}\mathbf{h}_i{-}\hat{\mathbf{h}}_i$ and
$\tilde{\mathbf{g}}_i{=}\mathbf{g}_i{-}\hat{\mathbf{g}}_i$. The
variances of the error vectors are
$\mathcal{E}\left[\parallel\tilde{\mathbf{h}}_i\parallel^2\right]{=}\sigma_{h,i}^2$
and
$\mathcal{E}\left[\parallel\tilde{\mathbf{g}}_i\parallel^2\right]{=}\sigma_{g,i}^2$.

\begin{figure}[t]
\renewcommand{\captionfont}{\small}
\captionstyle{center} \centering
\includegraphics[height=3cm,width=5cm]{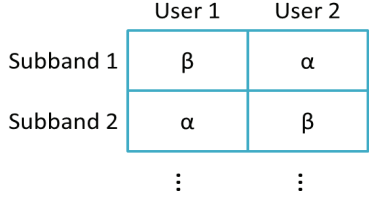}
\caption{The two-subband based scenario with imperfect CSIT.}\label{fig:scene}
\end{figure}

The CSIT setting in this two-subband based scenario is illustrated in Figure
\ref{fig:scene}.
 User 1 estimates its channel information in the first subband using pilots and feeds it back as
$\hat{\mathbf{h}}_1$ while user 2 reports its CSI in the second
subband as $\hat{\mathbf{g}}_2$. As shown, we assume the qualities of $\hat{\mathbf{h}}_1$ and $\hat{\mathbf{g}}_2$ are identical and expressed using a parameter, $\beta$, which is defined as
\begin{equation}
\beta \triangleq
{\lim}_{P\to\infty}{-}\frac{\log\sigma_{h,1}^2}{\log
P}={\lim}_{P\to\infty}{-}\frac{\log\sigma_{g,2}^2}{\log P}.
\end{equation}

As the CSI of two adjacent
subbands are correlated, the transmitter can predict the channel information of the unreported subband. To be specific, with the knowledge of
$\hat{\mathbf{h}}_1$, the channel condition of the second subband of user 1,
$\hat{\mathbf{h}}_2$, is predicted. Similarly, $\hat{\mathbf{g}}_1$ is predicted based on the knowledge of $\hat{\mathbf{g}}_2$. The qualities of these two predicted channel states are characterized as $\alpha$, which is defined as
\begin{equation}
\alpha \triangleq
{\lim}_{P\to\infty}{-}\frac{\log\sigma_{h,2}^2}{\log
P}={\lim}_{P\to\infty}{-}\frac{\log\sigma_{g,1}^2}{\log P}.
\end{equation}

\emph{Remarks:} 1) $\beta$ and $\alpha$ vary within the range of
$\left[0{,}1\right]$, where $0$ represents no CSIT whereas $1$
stands for perfect CSIT; 2) The quality of the predicted CSIT, $\hat{\mathbf{h}}_2$ and $\hat{\mathbf{g}}_1$, is bounded by the quality of $\hat{\mathbf{h}}_1$ and $\hat{\mathbf{g}}_2$, namely $\alpha{\leq}\beta$; 3) We assume that this two-subband scenario can be repeated an infinite number of times; 4) The transmitter and both users have the knowledge of $\hat{\mathbf{h}}_{1{:}2N}$ and
$\hat{\mathbf{g}}_{1{:}2N}$.
Besides, each receiver has perfect knowledge of local CSI; 5)  It is
important to note the quantities
$\mathcal{E}\left[|\mathbf{h}_1^H\hat{\mathbf{h}}_1^\bot|^2\right]{=}
\mathcal{E}\left[|\mathbf{g}_2^H\hat{\mathbf{g}}_2^\bot|^2\right]{\sim}
{P^{{-}\beta}}$ and
$\mathcal{E}\left[|\mathbf{g}_1^H\hat{\mathbf{g}}_1^\bot|^2\right]{=}
\mathcal{E}\left[|\mathbf{h}_2^H\hat{\mathbf{h}}_2^\bot|^2\right]{\sim}
{P^{{-}\alpha}}$.

Throughout the paper, we define a per-channel-use based \emph{DoF}, which is expressed as
\begin{equation}
d_i \triangleq \lim_{P\to\infty} \frac{R_i}{S\log P},\quad
i=1,2, \label{eq:dof_def}
\end{equation}
where $R_i$ is the rate achieved by user $i$ over $S$ channel uses\footnote{$S$ channel uses may physically refer to $S$ subbands with full transmission power $P$ or transmitting symbols using power of $P^{S}$ in a single subband. In this paper, it takes the latter understanding.}.

\section{\emph{DoF} Region with General CSIT Pattern}\label{DoF}

\newtheorem{mytheorem}{Theorem}
\begin{mytheorem} \label{DoF_theorem}
In a frequency correlated MISO BC with imperfect CSIT
shown in Figure \ref{fig:scene}, an achievable \emph{DoF} region is
characterized as a polygon composed of the following corner points:
\begin{multline}
\left\{\left(1{,}0\right){;}\left(0{,}1\right){;}\left(1{,}\alpha\right){;}\left(\alpha{,}1\right){;}{\cdots}\right.\\
\left.\left(\min\left(\frac{2{+}\alpha}{3}{,}\beta\right){,}
\max\left(\frac{2{+}\alpha}{3}{,}\frac{2{-}\beta{+}\alpha}{2}\right)\right){;}\cdots\right.\\\left.
\left(\max\left(\frac{2{+}\alpha}{3}{,}\frac{2{-}\beta{+}\alpha}{2}\right){,}
\min\left(\frac{2{+}\alpha}{3}{,}\beta\right)\right)\right\}.\label{eq:dof_reg}
\end{multline}
\end{mytheorem}

\begin{figure}[t]
\renewcommand{\captionfont}{\small}
\captionstyle{center} \centering
\includegraphics[height=6.5cm,width=8.7cm]{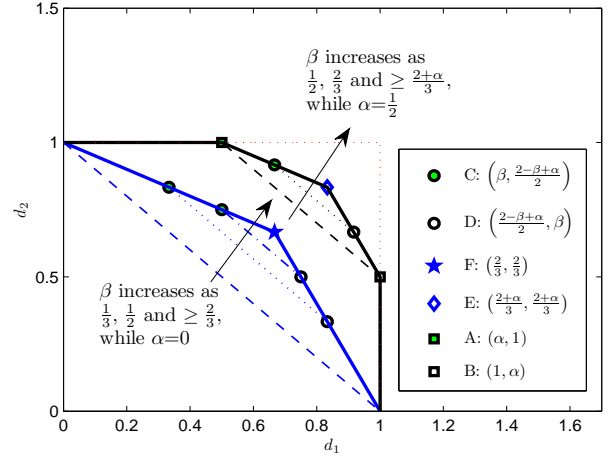}
\caption{Achievable \emph{DoF} region in frequency correlated
channel with imperfect CSIT.}\label{fig:dof}
\end{figure}

Figure \ref{fig:dof} illustrates the region specified by \eqref{eq:dof_reg}, spanning all $\beta$ and $\alpha$ satisfying $\beta{,}\alpha{\in}\left[0,1\right]$ and $\alpha{\leq}\beta$.
When $\alpha$ is fixed, points C and D (shown by circle points, see Figure \ref{fig:dof})
move closer to each other as $\beta$ increases. For $\beta{=}\frac{2{+}\alpha}{3}$,
points C and D join at point E (or F, see Figure
\ref{fig:dof}). If $\beta$ continues increasing, the \emph{DoF} region will not expand any further. This
reveals that the CSIT with quality $\beta$ satisfying
$\beta{\geq}\frac{2{+}\alpha}{3}$ can be as good as $\beta{=}1$.
Specifically, the \emph{DoF} region achieved by
MAT \cite{Tse10} with $\beta{=}1$ and $\alpha{=}0$ (composed of point F,
$\left(1{,}0\right)$ and $\left(0{,}1\right)$, see Figure \ref{fig:dof}) can be actually
achieved by $\beta{=}\frac{2}{3}$ and $\alpha{=}0$.

Moreover, if we fix $\beta$ and
increase $\alpha$, all the points will move either upwards or to the right. When $\alpha$ reaches $\beta$, point A will join point C while points B and D overlap. The \emph{DoF} region can be simply achieved by doing ZFBF plus superposition coding.

In addition, the maximum sum \emph{DoF} is achieved by the
diamond point E, and star point F, for the case
$\beta{\geq}\frac{2{+}\alpha}{3}$. Otherwise, it is obtained by the
circle points C and D.

\section{Achievability}\label{achievability}

\subsection{Motivations}

In this part, we briefly revisit two existing schemes, MAT and ZFBF. Their achievable rates in frequency correlated BC will be identified and analyzed. Their sub-optimalities will motivate the derivation of the novel transmission strategy.

\subsubsection{Reusing MAT scheme and Extensions}\label{MAT}

 In \cite{Tse10}, the transmission of MAT finishes in three time slots,
during which the transmit signal and received signals are
\begin{eqnarray}
\begin{array}{l}
\mathbf{s}_1=\mathbf{v}_1,\\
y_1=\eta_{1,1},\\
z_1=\mathbf{g}_1^H\mathbf{v}_1,
\end{array} &
\begin{array}{l}
\mathbf{s}_2=\mathbf{u}_2,\\
y_2=\mathbf{h}_2^H\mathbf{u}_2,\\
z_2=\eta_{2,2},
\end{array} &
\begin{array}{l}
\mathbf{s}_3=\left[\eta_{1,1}+\eta_{2,2},0\right]^T,\\
y_3=h_{3,1}^*\left(\eta_{1,1}+\eta_{2,2}\right),\\
z_3=g_{3,1}^*\left(\eta_{1,1}+\eta_{2,2}\right),
\end{array}\nonumber
\end{eqnarray}where $\eta_{1,1}{=}\mathbf{h}_1^H\mathbf{v}_1$ and
$\eta_{2,2}{=}\mathbf{g}_2^H\mathbf{u}_2$. User 1 receives its
desired symbol vector $\mathbf{u}_2$ in $y_2$ but overhears
$\mathbf{h}_1^H\mathbf{v}_1$ in $y_1$. The decoding is enabled once
the transmission at slot 3 is completed, where the sum of the
overheard interference is retransmitted. After decoding
$\eta_{1,1}{+}\eta_{2,2}$ received in $y_3$ and subtracting $y_1$, user 1 obtains an
additional independent observation of its desired symbol vector,
$\mathbf{g}_2^H\mathbf{u}_2$. Hence, user 1 can decode $\mathbf{u}_2$. Similarly, user
2 can decode $\mathbf{v}_1$. In this way, four symbols are successfully transmitted in three slots, resulting in the symmetric \emph{DoF} of $\frac{2}{3}$.

However, among all the six CSI in these three time slots, only two
of them are in fact employed, namely $\mathbf{h}_1^H$ and $\mathbf{g}_2^H$.
Equivalently, we can reuse MAT in the scenario shown in
Figure \ref{fig:scene} provided that $\beta{=}1$ and $\alpha{=}0$. The sum of the overheard
interference, $\eta_{1,1}{+}\eta_{2,2}$, is reconstructed and
retransmitted using an extra channel use (subband 3). The CSI of this extra channel use does not have to be known at the transmitter.

When $\beta{<}1$, the transmit power should be adjusted because the overheard interferences generated at each user are reconstructed with non-negligible error at the
transmitter. Specifically, after subtracting $y_1$ from
$y_3/h_{3,1}^*$, $\hat{\mathbf{g}}_2^H\mathbf{u}_2$ is obtained plus
a residue interference,
$\left(\hat{\mathbf{h}}_1^H{-}\mathbf{h}_1^H\right)\mathbf{v}_1{=}{-}\tilde{\mathbf{h}}_1^H\mathbf{v}_1$,
where
$\mathcal{E}\left[\parallel\tilde{\mathbf{h}}_1^H\parallel^2\right]{\sim}P^{{-}\beta}$.
To make the residue interference drowned by the noise, the
transmission power of $\mathbf{v}_1$ in subband 1 should be reduced
to $P^\beta$. In this way, $\beta$ channel use is
employed per subband, during which, both $\mathbf{v}_1$ and
$\mathbf{u}_2$ achieve the rate $2\beta{\log}P$ resulting in the sum
\emph{DoF} $\frac{4\beta}{3\beta}$ over $3\beta$ channel uses.

\subsubsection{Conventional approach-ZFBF}\label{ZFBF}

ZFBF is one of the conventional interference
mitigation techniques that achieve MU-MIMO transmission. The transmitter precodes two symbols $u_1$ and $v_1$ (intended to user 1 and 2 respectively) using the knowledge of CSIT of both users. The
transmission signal in subband 1 is expressed as
\begin{align}
\mathbf{s}_1 &= \hat{\mathbf{g}}_1^\bot u_1+\hat{\mathbf{h}}_1^\bot
v_1,\label{eq:ZFBFs}
\end{align}where $P_{u_1}{\sim}P^{\alpha}$ and $P_{v_1}{\sim}P^{\beta}$,
resulting in the received signals
\begin{align}
y_1 &= \mathbf{h}_1^H\hat{\mathbf{g}}_1^\bot
u_1+\mathbf{h}_1^H\hat{\mathbf{h}}_1^\bot
v_1+\epsilon_{1,y},\label{eq:ZFBFy}\\
z_1 &=\mathbf{g}_1^H\hat{\mathbf{g}}_1^\bot
u_1+\mathbf{g}_1^H\hat{\mathbf{h}}_1^\bot
v_1+\epsilon_{1,z}.\label{eq:ZFBFz}
\end{align}As the qualities of $\hat{\mathbf{h}}_1$ and $\hat{\mathbf{g}}_1$ are $\beta$ and $\alpha$ respectively,
$\mathbf{h}_1^H\hat{\mathbf{h}}_1^\bot v_1$ and
$\mathbf{g}_1^H\hat{\mathbf{g}}_1^\bot u_1$ are drowned by the noise.
In this way, the rate achieved by $u_1$ and $v_1$ are $\alpha{\log}P$ and
$\beta{\log}P$ respectively. The amount of channel use in subband 1 is
$\beta$. Similarly, the same transmission is applicable in subband 2
by switching the power of each user's symbol. Hence, the sum \emph{DoF} is $\frac{2\beta+2\alpha}{2\beta}$ during these $2\beta$ channel uses.

\subsubsection{Analysis and motivation}\label{Motivation}

\begin{table}[!h]
\renewcommand{\captionfont}{\small}
\captionstyle{center} \centering
\begin{tabular}{cccccc}
& Sum Rate  & Channel & $n_{1,1},n_{1,2}$ & $n_{2,1},n_{2,2}$ & Pre-\\
& ($\log P$) & Uses & & & coding\\
MAT & $4\beta$ & $3\beta$ & $0,2$ & $2,0$ & No\\
ZFBF & $2\beta{+}2\alpha$ & $2\beta$ & $1,1$ & $1,1$ & Yes\\
Objective & $4\beta$ & $<3\beta$ & $1,2$ & $2,1$ & Yes
\end{tabular}
\caption{Comparison among MAT, ZFBF and the objective. $n_{i,j}$
refers to the number of private symbols sent to user $i$ in subband $j$.}
\label{tab:MAT_ZFBF}
\end{table}

The comparison between MAT and ZFBF are presented in Table
\ref{tab:MAT_ZFBF}. ZFBF saves $\beta$ channel uses while
it incurs a rate loss of $2\left(\beta{-}\alpha\right){\log}P$ compared to MAT.

When $\alpha$ is small, MAT outperforms ZFBF in sum \emph{DoF}. In this case,
ZFBF precoding works inefficiently in rejecting the interference potentially seen by user 2 in subband 1, the transmit power of $u_1$ in \eqref{eq:ZFBFs} is therefore significantly limited, resulting in low \emph{DoF}. Similarly, user 2 achieves low rate in subband 2. However, MAT transmits two symbols to each user in turn. The CSIT with
quality $\beta$ is exploited to provide confident side information over an extra
channel use. The \emph{DoF} is therefore boosted up.

When $\alpha$ approaches $\beta$,  ZFBF works well in rejecting the interference potentially seen by both users in each subband. The sum rate achieved by ZFBF therefore approximates as $4\beta{\log}P$, resulting in a higher sum \emph{DoF} than MAT by saving the $\beta$ extra channel uses. However, MAT incurs a loss because the CSIT with quality $\alpha$ is wasted during the $3\beta$ channel uses.

Intuitively, given a certain value of $\alpha{\in}\left[0{,}\beta\right]$, a better sum \emph{DoF} can be obtained by a strategy
that optimally balances the employment of CSIT with quality $\alpha$
and the usage of extra channel use. This objective strategy can be designed as the
integration of ZFBF and MAT. It would outperform ZFBF by employing a
small fraction of extra channel use to perform overheard interference cancellation. At
the same time, as precoding is introduced, the amount of extra channel use could be reduced compared
to MAT while the sum rate remains $4\beta{\log}P$. The amount of extra channel use would be a function of $\beta$ and $\alpha$, bridging ZFBF and MAT. When $\alpha{=}\beta$, the transmission scheme would be upgraded to ZFBF; for $\alpha{=}0$, it would collapse to pure MAT. Bearing this in mind, we derive the novel
transmission block in the following section.

\subsection{Building New Transmission Blocks}\label{block}

Following the aforementioned motivation, the main features of
this strategy are presented in the last row of Table
\ref{tab:MAT_ZFBF}. It is a combination of ZFBF and MAT in terms of precoding and the number of symbols transmitted to each user per subband.

The transmission signals in subband 1 and 2 are respectively expressed as
\begin{align}
\mathbf{s}_1 &= \left[x_{c,1},0\right]^T{+}\left[\mu_1,0\right]^T{+}
\left[\hat{\mathbf{h}}_1^\bot,\hat{\mathbf{h}}_1\right]^T\mathbf{v}_1{+}\hat{\mathbf{g}}_1^\bot{u_1},\label{eq:s1}\\
\mathbf{s}_2 &= \left[x_{c,2},0\right]^T{+}\left[\mu_2,0\right]^T{+}
\left[\hat{\mathbf{g}}_2^\bot,\hat{\mathbf{g}}_2\right]^T\mathbf{u}_2{+}\hat{\mathbf{h}}_2^\bot{v_2},\label{eq:s2}
\end{align}
where two private symbols
($\mathbf{v}_1{=}\left[v_{1,1}{,}v_{1,2}\right]^T$) are transmitted
to user 2 and one private symbol ($u_1$) is sent to user 1 in subband 1.
Precoding is also considered in $\mathbf{s}_1$, where $v_{1,2}$ is
precoded with $\hat{\mathbf{h}}_1$, $v_{1,1}$ and $u_1$ are
projected to the orthogonal space of $\hat{\mathbf{h}}_1$ and
$\hat{\mathbf{g}}_1$ respectively. The new symbols in $\mathbf{s}_2$
are similarly encoded and transmitted.

Besides, $\mu_1$ and $\mu_2$ are two pieces of the quantized
overheard interference, encoded with the rates, $R_{\mu_1}{=}r_{\mu_1}{\log}P$ and
$R_{\mu_2}{=}r_{\mu_2}{\log}P$ respectively. $x_{c,1}$ and $x_{c,2}$ are common
messages that should be decoded by both users.
The power and rate allocation are presented in Table \ref{tab:pr}.
\begin{table}[!h]
\renewcommand{\captionfont}{\small}
\captionstyle{center} \centering
\begin{tabular}{c|ccc}
 & Symbols & Power & Encoding rate ($\log P$)\\
\hline
Subband 1 & $x_{c,1}$ & $P-P^{r_{\mu_1}+\beta}$ & $1-r_{\mu_1}-\beta$\\
& $\mu_1$ & $P^{r_{\mu_1}+\beta}-P^{\beta}$ & $r_{\mu_1}$\\
& $v_{1,1}$ & $P^{\beta}/2$ & $\beta$\\
& $v_{1,2}$ & $P^{\beta}/2-P^\alpha/2$ & $\beta-\alpha$\\
& $u_1$ & $P^\alpha/2$ &  $\alpha$\\
\hline
Subband 2 & $x_{c,2}$ & $P-P^{r_{\mu_2}+\beta}$ & $1-r_{\mu_2}-\beta$\\
& $\mu_2$ & $P^{r_{\mu_2}+\beta}-P^{\beta}$ & $r_{\mu_2}$\\
& $u_{2,1}$ & $P^{\beta}/2$ & $\beta$\\
& $u_{2,2}$ & $P^{\beta}/2-P^\alpha/2$ & $\beta-\alpha$\\
& $v_2$ & $P^\alpha/2$ &  $\alpha$
\end{tabular}
\caption{The power and rate allocated to the symbols in \eqref{eq:s1}
and \eqref{eq:s2}.} \label{tab:pr}
\end{table}

The received signals at each receiver in each subband are expressed
as
\begin{align}
y_1 &{=}
\underbrace{h_{1,1}^*x_{c,1}}_P{+}\underbrace{h_{1,1}^*\mu_1}_{P^{r_{\mu_1}{+}\beta}}{+}
\underbrace{\eta_{1,1}}_{P^\beta}{+}\underbrace{\mathbf{h}_1^H\hat{\mathbf{g}}_1^\bot{u_1}}_{P^\alpha}{+}\epsilon_{1,y},\label{eq:y1}\\
z_1 &{=}
\underbrace{g_{1,1}^*x_{c,1}}_P\!\!{+}\!\!\underbrace{g_{1,1}^*\mu_1}_{P^{r_{\mu_1}{+}\beta}}\!{+}\!
\underbrace{\mathbf{g}_1^H\left[\hat{\mathbf{h}}_1^\bot{,}\hat{\mathbf{h}}_1\right]\mathbf{v}_1}_{P^{\beta}}
\!{+}\!\underbrace{\mathbf{g}_1^H\hat{\mathbf{g}}_1^\bot{u_1}}_{P^0}\!{+}\epsilon_{1,z},\!\label{eq:z1}\\
y_2 &{=}
\underbrace{h_{2,1}^*x_{c,2}}_P\!\!{+}\!\!\underbrace{h_{2,1}^*\mu_2}_{P^{r_{\mu_2}{+}\beta}}\!{+}\!
\underbrace{\mathbf{h}_2^H\left[\hat{\mathbf{g}}_2^\bot{,}\hat{\mathbf{g}}_2\right]\mathbf{u}_2}_{P^\beta}
\!{+}\!\underbrace{\mathbf{h}_2^H\hat{\mathbf{h}}_2^\bot{v_2}}_{P^0}\!{+}\epsilon_{2,y},\!\label{eq:y2}\\
z_2 &{=}
\underbrace{g_{2,1}^*x_{c,2}}_P{+}\underbrace{g_{2,1}^*\mu_2}_{P^{r_{\mu_2}{+}\beta}}{+}
\underbrace{\eta_{2,2}}_{P^\beta}{+}\underbrace{\mathbf{g}_2^H\hat{\mathbf{h}}_2^\bot{v_2}}_{P^\alpha}{+}\epsilon_{2,z},\label{eq:z2}
\end{align}where $\eta_{1,1}$ and $\eta_{2,2}$ are the overheard
interferences generated at user 1 in subband 1 and at user 2 in
subband 2 respectively,
\begin{align}
\eta_{1,1}&{=}
\underbrace{\mathbf{h}_1^H\hat{\mathbf{h}}_1^\bot{v_{1,1}}}_{P^0}{+}
\underbrace{\mathbf{h}_1^H\hat{\mathbf{h}}_1{v_{1,2}}}_{P^\beta},\label{eq:eta1}\\
\eta_{2,2}&{=}
\underbrace{\mathbf{g}_2^H\hat{\mathbf{g}}_2^\bot{u_{2,1}}}_{P^0}{+}
\underbrace{\mathbf{g}_2^H\hat{\mathbf{g}}_2{u_{2,2}}}_{P^\beta}.\label{eq:eta2}
\end{align}Note that the power stated below each term is obtained asymptotically, which is merely valid at high SNR. Since
$\mathcal{E}\left[|\mathbf{\mathbf{h}}_1^H\hat{\mathbf{h}}_1^\bot|^2\right]{\sim}P^{{-}\beta}$,
the term $\mathbf{h}_1^H\hat{\mathbf{h}}_1^\bot{v_{1,1}}$ in
$\eta_{1,1}$ is drowned by the noise in \eqref{eq:y1}. Similarly,
$\mathbf{g}_2^H\hat{\mathbf{g}}_2^\bot{u_{2,1}}$ in $\eta_{2,2}$
vanishes.

Hence, the overheard interference $\eta_{1,1}$ and $\eta_{2,2}$ are
composed of $v_{1,2}$ and $u_{2,2}$ respectively, which are then
possible to be detected at receiver 1 and 2 respectively (will be
discussed in Section \ref{cd}). In this way, when reconstructing the
sum of the overheard interferences, the channel component, for
instance, $\mathbf{h}_1^H\hat{\mathbf{h}}_1$ in \eqref{eq:eta1}, can
be dropped. As a consequence, in contrast to MAT where
$\eta_{1,1}{+}\eta_{2,2}$ is rebuilt and sent, we reconstruct the
sum of the symbols $v_{1,2}$ and $u_{2,2}$ as
\begin{equation}
\mu = v_{1,2}+u_{2,2}.\label{eq:mu}
\end{equation}
$\mu$ can be generated from a codebook $\left\{v_{1,2}{+}u_{2,2}\right\}$. Since $v_{1,2}$ and $u_{2,2}$ are encoded with same rate, we assume they are generated from the same codebook, denoted as $\Phi$. Moreover, we design $\Phi$ as a set close to the arithmetic plus\footnote{$\Phi$ close to arithmetic plus is defined as, if $a{\in}\Phi$ and $b{\in}\Phi$, then $a{+}b{\in}\Phi$.}. In this way, $\mu{=}v_{1,2}{+}u_{2,2}$ can be generated from $\Phi$ as well, with the encoding rate, $r_\mu{=}\beta{-}\alpha$, identical to that for $v_{1,2}$ and $u_{2,2}$

Furthermore, as motivated by \cite{Chen12a} and \cite{Chen12b}, we split $\mu$ into two parts, $\mu_1$ and $\mu_2$ as
\begin{equation}
\mu\triangleq\left\{\mu_1,\mu_2\right\},
\end{equation}
each is encoded from a codebook $\Phi_1$ and $\Phi_2$ respectively, which are independent to each other. The encoding rates of these two codebooks are subject to
\begin{align}
R_{\mu_1}+R_{\mu_2}&\approx\left(\beta-\alpha\right)\log P,\\
r_{\mu_1}+r_{\mu_2}&=\beta-\alpha.\label{eq:mu1_mu2}
\end{align}
Hence, $\Phi$ can be considered as a product set of $\Phi_1$ and $\Phi_2$. $\mu_1$ and $\mu_2$ are decoded separately in two parallel channels, $\mu$ can therefore be perfectly reconstructed by combining them.

As presented in Table \ref{tab:pr}, $\mu_1$ and $\mu_2$ are respectively superposed on the private
symbols transmitted in subband 1 and 2. However, their power $P^{r_{\mu_1}{+}\beta}$ and
$P^{r_{\mu_2}{+}\beta}$, should not exceed the power constraint $P$. This constraint
can be expressed as
\begin{equation}
r_{\mu_1}\leq1-\beta,\quad r_{\mu_2}\leq1-\beta. \label{eq:rmu12}
\end{equation}
As a consequence, the transmission is subject to the relationship between
\eqref{eq:mu1_mu2} and \eqref{eq:rmu12}.

First, in the case of $2\left(1{-}\beta\right){>}\beta{-}\alpha$,
namely $\beta{<}\frac{2{+}\alpha}{3}$, the power of
$\mu_1$ and $\mu_2$ does not exceed the per-subband power
constraint by simply setting $r_{\mu_1}{=}r_{\mu_2}{=}\frac{\beta{-}\alpha}{2}$. Moreover, we can superimpose a common message $x_{c,1}$
on $\mu_1$ in subband 1 and $x_{c,2}$ on $\mu_2$ in subband 2 using
power stated in Table \ref{tab:pr}, which is scaled with $P$. Second, when
$2\left(1{-}\beta\right){=}\beta{-}\alpha$, the power constraint is
still satisfied but no common messages is transmitted since
$r_{\mu_1}{+}\beta{=}1$. Third, for the case of
$2\left(1{-}\beta\right){<}\beta{-}\alpha$, namely
$\beta{>}\frac{2{+}\alpha}{3}$, the value of $r_{\mu_1}$ and
$r_{\mu_2}$ are bounded by $1{-}\beta$ as in \eqref{eq:rmu12}.
Therefore, $\mu$ has to be divided into three pieces as
\begin{equation}
\hat{\mu}\triangleq\left\{\mu_1,\mu_2,\mu_3\right\},
\end{equation}
with the rates given by
\begin{eqnarray}
&R_{\mu_1}=R_{\mu_2}\approx\log P^{r_{\mu_1}},\quad R_{\mu_3}\approx\log P^{r_{\mu_3}},&\\
&r_{\mu_1}=r_{\mu_2}=1-\beta,\quad r_{\mu_3}=3\beta-\alpha-2.&\label{eq:mu123}
\end{eqnarray}
The transmission of $\mu_3$ requires an extra channel use in another subband. Next, we will discuss the achievabilities of each point in Figure \ref{fig:dof} depending on the requirement of extra channel use.

\subsection{Case I: $\beta{\leq}\frac{2{+}\alpha}{3}$-Achieving Points C and
D}\label{cd}

In this case, $\mu$ is split into two parts and no extra channel use is required. Messages $x_{c,1}$ and $x_{c,2}$ are transmitted provided that $\beta{<}\frac{2{+}\alpha}{3}$. The decoding procedure
is described as follows.

\subsubsection{\textbf{Stage I}-Decode $x_{c,1}$ and $x_{c,2}$}\label{xc}

Revisiting \eqref{eq:y1} and $\eqref{eq:z1}$, the received power of
$x_{c,1}$ is $P$ at each receiver. Successive interference
cancelation (SIC) is selected as the decoding strategy. $x_{c,1}$ is
decoded at the first stage treating all the other symbols as noise.
Consequently, the rates of $x_{c,1}$ achieved at user 1 and user 2
are
$R_{x_{c,1}}^{\left(1\right)}{=}I\left(x_{c,1};y_1|\mathbf{h}_1\right)$
and
$R_{x_{c,1}}^{\left(2\right)}{=}I\left(x_{c,1};z_1|\mathbf{g}_1\right)$,
respectively. These two rates are equal to
${\log}\frac{P{-}P^{r_{\mu_1}{+}\beta}}{P^{r_{\mu_1}{+}\beta}}$,
which is asymptotically $\left(1{-}r_{\mu_1}{-}\beta\right){\log}P$
for infinite $P$. Similarly, $x_{c,2}$ achieves the rate
$\left(1{-}r_{\mu_2}{-}\beta\right){\log}P$ in $y_2$ and $z_2$.

\subsubsection{\textbf{Stage II}-Decode $\mu_1$, $\mu_2$ and obtain
$\hat{\mu}$}\label{dec_eta}

As $\mu_1$ and $\mu_2$ are independently encoded and
sent in subband 1 and 2 respectively, they can be decoded separately at both receivers. After that, $\mu$ is obtained by combining them.

In $y_1$ and $z_1$, $\mu_1$ is decoded at the second stage of SIC,
where $x_{c,1}$ has been completely subtracted. Treating all the
component to the r.h.s. of $\mu_1$ in \eqref{eq:y1} and
\eqref{eq:z1} as noise, $\mu_1$ is decoded with the rate of
$R_{\mu_1}^{\left(1\right)}{=}I\left(\mu_1;y_1|\mathbf{h}_1,x_{c,1}\right)$
and
$R_{\mu_1}^{\left(2\right)}{=}I\left(\mu_1;z_1|\mathbf{g}_1,x_{c,1}\right)$
by user 1 and user 2 respectively. Both quantities are
equal to ${\log}\frac{P^{r_{\mu_1}{+}\beta}{-}P^\beta}{P^\beta}$,
which is $r_{\mu_1}\log P$ at high SNR. Similarly, $\mu_2$ is
decoded with rate $r_{\mu_2}\log P$ in subband 2. After that,
$\mu_1$ and $\mu_2$ have been successfully decoded so that
$\mu$ is completely recovered.

\subsubsection{\textbf{Stage III}-Decode $u_1$ and $v_2$}\label{dec_sym}

Employing SIC as the decoding strategy, $u_1$ and $v_2$ are decoded from $y_1$ and $z_2$ respectively.

Let us introduce a notation, $y_1^\prime$, representing the signal after subtracting $x_{c,1}$ and $\mu_1$ as
\begin{align}
y_1^\prime&=y_1-h_{1,1}^*\left(x_{c,1}+\mu_1\right)\\
&=\eta_{1,1}+\mathbf{h}_1^H\hat{\mathbf{g}}_1^\bot{u_1}+\epsilon_{1,y}\\
&=\mathbf{h}_1^H\hat{\mathbf{h}}_1v_{1,2}+\mathbf{h}_1^H\hat{\mathbf{g}}_1^\bot{u_1}+\epsilon_{1,y}^\prime,\label{eq:y1p}
\end{align}
where $\eta_{1,1}$ is given in \eqref{eq:eta1} and
$\epsilon_{1,y}^\prime$ results from merging $\epsilon_{1,y}$ and
$\mathbf{h}_1^H\hat{\mathbf{h}}_1^\bot v_{1,1}$.
Treating $\mathbf{h}_1^H\hat{\mathbf{g}}_1^\bot{u_1}$ as noise, $v_{1,2}$ can be decoded
with the rate
$R_{v_{1,2}}^{\left(1\right)}{=}I\left(v_{1,2};y_1^\prime|\mathbf{h}_1,\hat{\mathbf{h}}_1\right){=}{\log}\frac{P^{\beta}{-}P^\alpha}{P^\alpha}$,
which is asymptotically equal to $\left(\beta{-}\alpha\right)\log P$
at high SNR. After that, $u_1$ is seen by subtracting
$\mathbf{h}_1^H\hat{\mathbf{h}}_1v_{1,2}$ from $y_1^\prime$ as
\begin{align}
y_1^{\prime\prime}=y_1^\prime-\mathbf{h}_1^H\hat{\mathbf{h}}_1v_{1,2}
&=\mathbf{h}_1^H\hat{\mathbf{g}}_1^\bot{u_1}+\epsilon_{1,y}^\prime.
\label{eq:y1pp}
\end{align}
The rate of $u_1$ is $R_{u_1}^{\left(1\right)}{=}
I\left(u_1{;}y_1^{\prime\prime}|\mathbf{h}_1{,}\hat{\mathbf{g}}_1\right){=}\alpha{\log}P$.
Similarly, employing SIC to $z_2$ results in
$R_{v_2}^{\left(2\right)}{=}\alpha{\log}P$ and
$R_{u_{2,2}}^{\left(2\right)}{=}\left(\beta{-}\alpha\right){\log}P$.

\subsubsection{\textbf{Stage IV}-Decode $\mathbf{v}_1$ and $\mathbf{u}_2$}

As $\mu$ has been recovered perfectly in Stage II, each user can get access to $v_{1,2}{+}u_{2,2}$. From \eqref{eq:mu}, the rate of $u_{2,2}$ is obtained as
$R_{u_{2,2}}^{\left(1\right)}{=}I\left(u_{2,2};\mu|v_{1,2}\right)$,
where having the knowledge of $v_{1,2}$ is the prerequisite. As
$v_{1,2}$ was successfully decoded at user 1 in Stage III, it can be
completely removed from \eqref{eq:mu}, resulting in
$R_{u_{2,2}}^{\left(1\right)}{=}\left(\beta{-}\alpha\right){\log}P$.
Similarly the rate of $v_{1,2}$ at receiver 2 is
$R_{v_{1,2}}^{\left(2\right)}{=}I\left(v_{1,2};\mu|u_{2,2}\right){=}\left(\beta{-}\alpha\right){\log}P$.

After decoding $u_{2,2}$, $u_{2,1}$ is decodable from $y_2$.
Denoting $y_2^\prime{=}y_2{-}h_{2,1}^*\left(x_{c,2}{+}\mu_2\right)$, merging $\mathbf{h}_2^H\hat{\mathbf{h}}_2^\bot{v_2}$ and the
noise $\epsilon_{2,y}$ into $\epsilon_{2,y}^\prime$, we have
\begin{align}
y_2^\prime{=}\mathbf{h}_2^H\hat{\mathbf{g}}_2^\bot{u_{2,1}}+\mathbf{h}_2^H\hat{\mathbf{g}}_2u_{2,2}+\epsilon_{2,y}^\prime.\label{eq:y2p}
\end{align}
$u_{2,1}$ is obtained by removing
$\mathbf{h}_2^H\hat{\mathbf{g}}_2u_{2,2}$ from $y_2^\prime$,
resulting in the rate
$R_{u_{2,1}}^{\left(1\right)}{=}I\left(u_{2,1};y_2^\prime|u_{2,2},\mathbf{h}_2,\hat{\mathbf{g}}_2\right){=}\beta{\log}P$.
Similarly, $v_{1,1}$ is decoded with the rate
$R_{v_{1,1}}^{\left(2\right)}{=}\beta{\log}P$.

To sum up, the \emph{DoF} achieved in these two subbands are
\begin{align}
d_1 {=}&\lim_{P\to\infty}\!\!\!
\frac{R_{x_{c,1}}^{\left(1\right)}{+}R_{x_{c,2}}^{\left(1\right)}{+}R_{u_1}^{\left(1\right)}{+}
R_{u_{2,1}}^{\left(1\right)}{+}R_{u_{2,2}}^{\left(1\right)}}{2\log P}{=}\frac{2{+}\alpha{-}\beta}{2},\label{eq:d1cd}\\
d_2 {=}&\lim_{P\to\infty} \frac{R_{v_{1,1}}^{\left(2\right)}{+}
R_{v_{1,2}}^{\left(2\right)}{+}R_{v_2}^{\left(2\right)}}{2\log P}{=}\beta,\label{eq:d2cd}
\end{align}
where we assume $x_{c,1}$ and $x_{c,2}$ are intended to user 1 so
that point D is achieved. Similarly, point C is achieved if
$x_{c,1}$ and $x_{c,2}$ are intended to user 2.

\subsection{Case II: $\beta{\geq}\frac{2{-}2\alpha}{3}$-Achieving Point
E}\label{e}

In this case, we remind the reader of the discussion in Section
\ref{block} that $\mu$ is split into three parts and an extra channel use is required to transmit
$\mu_3$. Besides, no common message is transmitted.

To achieve point E, we
repeat the transmission blocks in \eqref{eq:s1} and \eqref{eq:s2}
for $L$ times and employ one additional subband, namely subband
$2L{+}1$, to finalize the transmissions of
$\mu_{3,i}{,}i{=}1{,}2{,}{\cdots}{,}L$, where $\mu_{3,i}$ refers to
the third piece of overheard interference generated in subband $2i{-}1$ and $2i$. The rate of $\mu_{3,i}$ is denoted as $r_{\mu_{3,i}}{\log}P$ and we assume $r_{\mu_{3,i}}{=}r_{\mu_3}{,}i{=}1{,}2{,}{\cdots}{,}L$. The
quality of CSIT in subband $2L{+}1$ is identical to subband 1.

The transmission in subband $2L{+}1$ is expressed as
\begin{multline}
s_{2L+1} {=} \left[\mu_{3,1}{,}
0\right]^T{+}\left[\mu_{3,2}{,}0\right]^T{+}{\cdots}{+}\left[\mu_{3,L}{,}0\right]^T{+}\cdots\\
\hat{\mathbf{g}}_{2L{+}1}^\bot
u_{2L{+}1}{+}\hat{\mathbf{h}}_{2L{+}1}^\bot{v_{2L{+}1}},\label{eq:2L+1}
\end{multline}with the power and rate allocation presented in
Table \ref{tab:prL}.
\begin{table}[!h]
\renewcommand{\captionfont}{\small}
\captionstyle{center} \centering
\begin{tabular}{ccc}
Symbols & Power & Encoding rate \\
$\mu_{3,1}$ & $P-P^{1-r_{\mu_3}}$ & $r_{\mu_3}$\\
$\mu_{3,2}$ & $P^{1-r_{\mu_3}}-P^{1-2r_{\mu_3}}$ & $r_{\mu_3}$\\
$\mu_{3,3}$ & $P^{1-2r_{\mu_3}}-P^{1-3r_{\mu_3}}$ & $r_{\mu_3}$\\
$\vdots$ & $\vdots$ & $\vdots$\\
$\mu_{3,L}$ & $P^{1-\left(L-1\right)r_{\mu_3}}-P^{\alpha}$ & $r_{\mu_3}$\\
$u_{2L+1}$ & $P^{\alpha}/2$ & $\alpha$\\
$v_{2L+1}$ & $P^{\alpha}/2$ & $\alpha$
\end{tabular}
\caption{Power and rate allocation in subband $2L{+}1$ in case II.}
\label{tab:prL}
\end{table}Considering $\mathbf{s}_{2L{+}1}$ and the transmit power,
the received signal at user 1 is given as
\begin{multline}
y_{2L+1}{=}h_{2L{+}1,1}^*\left(\right.\!\!\underbrace{\mu_{3,1}}_P\!\!{+}\!\!\underbrace{\mu_{3,2}}_{P^{1{-}r_{\mu_3}}}\!\!{+}\!\!
\underbrace{\mu_{3,3}}_{P^{1{-}2r_{\mu_3}}}\!\!\!\!{+}\cdots{+}\!\!\!\!\!
\underbrace{\mu_{3,L}}_{P^{1-\left(L-1\right)r_{\mu_3}}}\!\!\!\!\!\!\!\left.\right){+}\cdots\\
\underbrace{\mathbf{h}_{2L{+}1}^H\hat{\mathbf{g}}_{2L{+}1}^\bot{u_{2L{+}1}}}_{P^\alpha}{+}
\underbrace{\mathbf{h}_{2L{+}1}^H\hat{\mathbf{h}}_{2L{+}1}^\bot{v_{2L{+}1}}}_{P^0}{+}\epsilon_{2L{+}1,y},\label{eq:2L+1}
\end{multline}
where all the symbols are decodable using SIC. Specifically, after
$\mu_{3{,}1{:}i{-}1}$ are decoded, $\mu_{3{,}i}$ are decoded
treating all the components to the r.h.s. of it in \eqref{eq:2L+1}
as noise. The rate achieved for $\mu_{3,i}{,}i{<}L$ is derived as
\begin{align}
R_{\mu_{3,i}}^{\left(1\right)}& {=}I\left(\mu_{3,i};y_{2L{+}1}|\mathbf{h}_{2L+1},
\mu_{3,1:i{-}1}\right)\\
& {=}{\log}\frac{P_{\mu_{3,i}}}{P_{u_{2L+1}}+\sum_{j=i+1}^L P_{\mu_{3,j}}}\\
& {=}\log \frac{P^{1-\left(i-1\right)r_{\mu_3}}-P^{1-ir_{\mu_3}}}{P^{1-ir_{\mu_3}}}\approx r_{\mu_3}\log P.
\end{align}

After decoding $\mu_{3{,}1{:}L{-}1}$, $\mu_{3{,}L}$ can be decoded treating $u_{2L{+}1}$ as noise. The rate of $\mu_{3{,}L}$ is
$R_{\mu_{3{,}L}}^{\left(1\right)}{\approx}{\log}\frac{P^{1{-}\left(L{-}1\right)r_{\mu_3}}}{P^\alpha}$,
whose pre-log factor is $1{-}\left(L{-}1\right)r_{\mu_3}{-}\alpha$.
To make $\mu_{3{,}L}$ decodable with rate $r_{\mu_3}$, $L$ should
satisfy the condition
$1{-}\left(L{-}1\right)r_{\mu_3}{-}\alpha{=}r_{\mu_3}$, resulting in
\begin{equation}
L{=}\frac{1{-}\alpha}{r_{\mu_3}}{=}\frac{1{-}\alpha}{3\beta{-}\alpha{-}2}. \label{eq:L}
\end{equation}
Similarly, user 2 can decode
$\mu_{3{,}1{:}L}$ using SIC.

Consequently, $\hat{\mu}_{1{:}L}$ can be recovered by collecting and
combining $\mu_{1{,}1{:}L}$, $\mu_{2{,}1{:}L}$ and
$\mu_{3{,}1{:}L}$. Moreover, all the symbols transmitted from the
1st to $2L$th subband are decodable using the decoding flow described
in Section \ref{cd}. The rates achieved in subband $2i{-}1$ and $2i$
are
$R_{u_{2i{-}1}}^{\left(1\right)}{=}R_{v_{2i}}^{\left(2\right)}{=}\alpha{\log}P$,
$R_{u_{2i,1}}^{\left(1\right)}{=}R_{v_{2i{-}1,1}}^{\left(2\right)}{=}\beta{\log}P$
and
$R_{u_{2i,2}}^{\left(1\right)}{=}R_{v_{2i{-}1,2}}^{\left(2\right)}{=}\left(\beta{-}\alpha\right){\log}P$.

Besides, $u_{2L{+}1}$ is decoded in \eqref{eq:2L+1} with rate
$R_{u_{2L{+}1}}^{\left(1\right)}{=}\alpha{\log}P$ at the last stage
of SIC after removing all the $\mu_{3{,}i}$. Similarly,
$R_{v_{2L{+}1}}^{\left(2\right)}{=}\alpha{\log}P$ is achieved at
receiver 2. Finally, we can conclude the \emph{DoF} achieved by each
user as
\begin{align}
d_1 \!&{=}\!\!\lim_{P\to\infty}\!\!
\frac{L{\times}\!\left(\!R_{u_{2i{-}1}}^{\left(1\right)}{+}R_{u_{2i,1}}^{\left(1\right)}
{+}R_{u_{2i,2}}^{\left(1\right)}\right)\!\!{+}R_{u_{2L{+}1}}^{\left(1\right)}}{\left(2L{+}1\right)\log P}{=}\frac{2{+}\alpha}{3},\label{eq:d1E}\\
d_2 \!&{=}\!\!\lim_{P\to\infty}\!\!
\frac{L{\times}\!\left(\!R_{v_{2i}}^{\left(2\right)}{+}R_{v_{2i{-}1,1}}^{\left(2\right)}\!
{+}R_{v_{2i{-}1,2}}^{\left(2\right)}\!\right)\!\!{+}R_{v_{2L{+}1}}^{\left(2\right)}}{\left(2L{+}1\right)\log P}{=}\frac{2{+}\alpha}{3}.\!\label{eq:d2E}
\end{align}

In the proposed scheme, the transmissions of private symbols in subband
1 and 2 occupy $2\beta$ channel uses while transmitting $\hat{\mu}$
requires $\beta{-}\alpha$ channel uses. Over those channel uses, the
sum rate of the private symbols are $4\beta{\log}P$, resulting in the
sum \emph{DoF} $\frac{4\beta}{3\beta{-}\alpha}$. Revisiting Table
\ref{tab:MAT_ZFBF}, our new scheme achieves the same sum rate as MAT
but using less channel uses. At the same time, it outperforms ZFBF
by $2\left(\beta{-}\alpha\right){\log}P$ in sum rate with only
$\beta{-}\alpha$ more channel use.

\subsection{''SC+ZF''-Achieving Points A and B}

Points A and B can be simply achieved via ZFBF using $\alpha$
channel use and transmitting common message $x_{c,i}$ using
the remaining $1{-}\alpha$ channel use in each subband. The
transmitted signal in subband 1 is expressed as
\begin{align}
\mathbf{s}_1 & =\left[x_{c,1},0\right]^T+\hat{\mathbf{g}}_1^\bot
u_1+\hat{\mathbf{h}}_1^\bot v_1,\label{eq:sczf}
\end{align}where $u_1$, $v_1$ are allocated with power $P^{\alpha}/2$
and the power of $x_{c,1}$ is $P{-}P^\alpha$. User 1 observes
$x_{c,1}$ plus $\mathbf{h}_1^H\hat{\mathbf{g}}_1^\bot{u_1}$ because
$\mathbf{h}_1^H\hat{\mathbf{h}}_1^\bot{v_1}$ is drowned by the noise.
Treating $\mathbf{h}_1^H\hat{\mathbf{g}}_1^\bot u_1$ as noise, $x_{c,1}$ is decoded at the first stage of SIC with rate
$\left(1{-}\alpha\right)\log P$. After
that, $u_1$ is decoded with the rate $\alpha$. Similarly, user 2 can
decode $x_{c,1}$ and $v_1$. The transmission and decoding procedure
are applicable to subband 2. As a result, assuming $x_{c,1}$ and
$x_{c,2}$ are intended to user 1, the \emph{DoF} are expressed as
\begin{align}
d_1 &{=} \lim_{P\to\infty}
\frac{R_{x_{c,1}}^{\left(1\right)}{+}R_{x_{c,2}}^{\left(1\right)}{+}R_{u_1}^{\left(1\right)}{+}R_{u_2}^{\left(1\right)}}{2\log P}{=}1,\\
d_2 &
{=}\lim_{P\to\infty}\frac{R_{v_1}^{\left(2\right)}{+}R_{v_2}^{\left(2\right)}}{2\log P}{=}\alpha,\label{eq:dofab}
\end{align}
so that point B is achieved. Point A is achieved if the common messages are assumed intended to user 2.

\section{Conclusion}\label{conclusions}

This work for the first time in the literature investigates the
impact of imperfect CSIT on the \emph{DoF} region of frequency
correlated MISO BC. A general two-subband based imperfect CSIT
pattern (see Figure \ref{fig:scene}) is studied. MAT and ZFBF achieve the optimal sum \emph{DoF} for $\alpha{=}0$ and $\alpha{=}\beta$ respectively while both of them have \emph{DoF} loss for $0{<}\alpha{<}\beta$. A novel transmission strategy is proposed to improve the performance. It processes as an integration of MAT and ZFBF, where precoding and overheard interference cancellation are combined. The \emph{DoF} region achieved is a function of $\beta$ and $\alpha$, enhancing the result for $0{<}\alpha{<}\beta$ and bridging the region of MAT (for $\alpha{=}0$) and ZFBF (for $\alpha{=}\beta$).

More details on the achievability and the converse will be provided in our upcoming journal version paper. Besides, we will investigate a more general scenario where the qualities of the four CSIT in Figure \ref{fig:scene} are all different. The proposed scheme and the \emph{DoF} region will be extended.

\bibliographystyle{IEEEtran}

\bibliography{icc13freqbib}

\end{document}